\documentclass{appolb}
\usepackage{epsfig}
\usepackage{graphicx}
\begin{document}
\title{
Polarization of Prompt $J/\psi$ and  $\Upsilon(nS)$
\thanks{Presented at 
the X International Workshop on Deep Inelastic Scattering (DIS2002),
Cracow, Poland, 30 April - 4 May 2002.
}%
}
\author{Jungil Lee 
\address{HEP Division, Argonne National Laboratory, 
\\9700 S. Cass Avenue, Argonne, Illinois 60439, U.S.A.}
}
\maketitle
\begin{abstract}
We review predictions, based on the nonrelativistic QCD factorization
framework, for the polarizations of prompt $J/\psi$'s  and $\Upsilon(nS)$'s
produced at the Fermilab Tevatron.  We also discuss the effect of 
relativistic corrections on the theoretical prediction for the
polarization of prompt $J/\psi$'s at the Tevatron. 
\end{abstract}
\PACS{13.85.Ni, 13.88.+e, 12.38.Bx, 13.87.Fh, 14.80.Ly}
\section{Introduction}
\vspace{-10cm}\hspace{8cm}{ANL-HEP-CP-02-045}

\vspace{9.8cm}
The nonrelativistic QCD~(NRQCD) factorization approach has been developed
to describe inclusive quarkonium production and decay~\cite{BBL}. 
It is the first effective field theory providing infrared-finite predictions 
of the hadronic decay rates of $P$-wave quarkonium~\cite{BBL-p}. 
It also explains 
large-$p_T$ $S$-wave charmonium production at the Tevatron~\cite{BF-frag}.
The theory introduces several nonperturbative color-octet 
matrix elements~(ME's). It also includes color-singlet ME's.
These are equivalent to the heavy-quark wave function at the origin
and it's derivatives, which appear in the color-singlet model. 
The ME's are universal and fitted
to the CDF data at the Tevatron~\cite{cho}. 
The universality of the ME's has been tested in various experimental 
situations~\cite{test}(A recent review can be found in Chap. 9
of Ref.~\cite{RUNII}).
A remarkable prediction of the NRQCD is that the
$S$-wave quarkonium produced in the $p\bar{p}$ collision
should be transversely polarized at sufficiently large $p_T$~\cite{CW}. 
This prediction is based on the dominance of gluon fragmentation 
into quarkonium in quarkonium
production at large $p_T$~\cite{BF-frag} and on the approximate heavy-quark 
spin symmetry of NRQCD~\cite{BBL}.  
Recent measurements at the Tevatron by the CDF Collaboration do not confirm
this prediction~\cite{psipol-cdf}.
However, the experimental uncertainties are large.

In this proceedings, we review predictions, based on the NRQCD 
factorization framework, for the polarizations of prompt 
$J/\psi$'s~\cite{psipol} and $\Upsilon(nS)$'s 
produced at the Tevatron~\cite{upspol}. 
As a possible source of depolarizing contributions to the prompt
$J/\psi$ production at large $p_T$, 
we consider the relativistic corrections to the gluon fragmentation 
into $J/\psi$. Finally, we summarize recent developments in the 
application of NRQCD to low-energy supersymmetry.
\section{Polarization of prompt $J/\psi$}
\begin{figure}
\begin{tabular}{ll}
\raisebox{2.05ex}{\includegraphics[width=6cm]{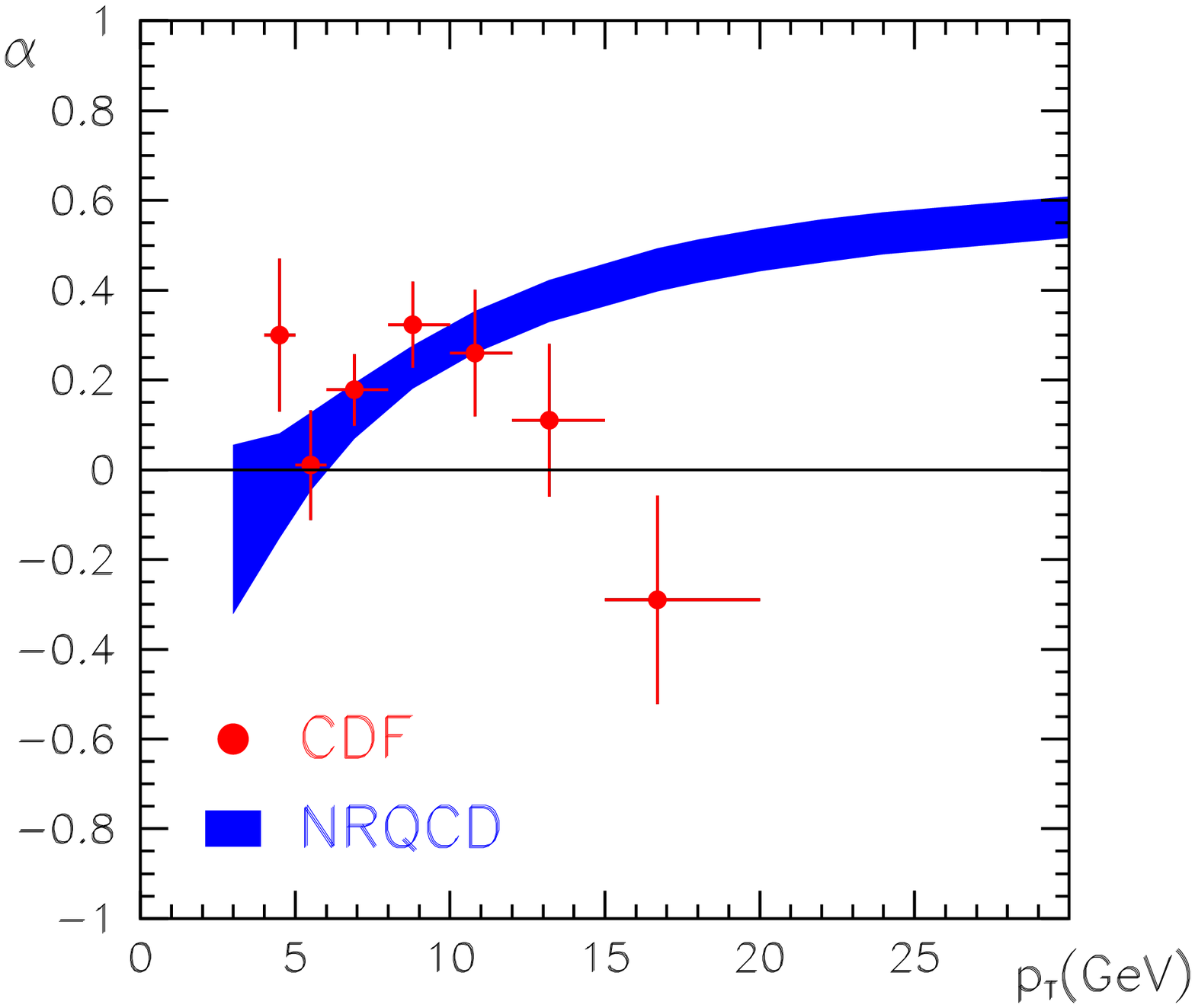}}&
\includegraphics[width=5.6cm]{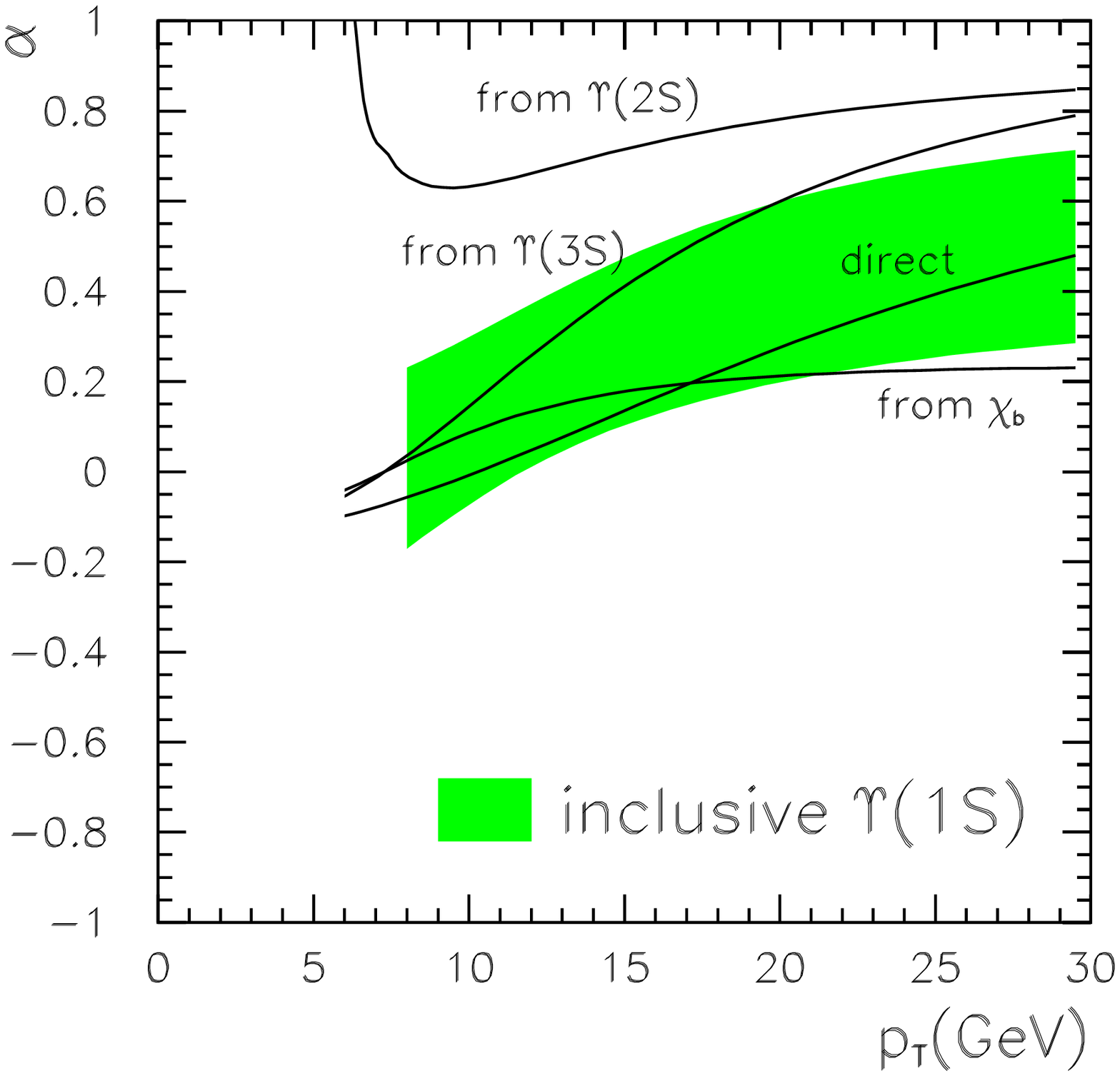}
\\[-5cm]
\hspace{1.2cm}(a)&
\hspace{1.2cm}(b)
\\[5cm]
\end{tabular}
\caption{Polarization variable $\alpha$ vs. $p_T$
(a) for prompt $J/\psi$~\cite{psipol} and (b) $\Upsilon(1S)$~\cite{upspol}
compared to CDF data.}
\end{figure}
A convenient measure of the polarization is the variable
$\alpha = (\sigma_T-2\sigma_L)/(\sigma_T+2\sigma_L)$,
where $\sigma_T$ and $\sigma_L$  are the transverse
and longitudinal components of the cross section, respectively.
The variable $\alpha$ is fitted to the angular distribution~(%
$\propto 1 + \alpha \cos^2\theta$) 
of the positive lepton with respect to the
$J/\psi$ momentum in the hadron center-of-momentum frame.
The polarizations of the $\psi'$'s and direct $J/\psi$'s~($J/\psi$'s that 
do not come from decays) produced
at the Tevatron are predicted to be transverse~\cite{psi2pol,psipol}.
The CDF measurement does not show the predicted tendency, but the
error bars are too large to draw any definitive conclusions~\cite{psipol-cdf}. 
CDF also measured the polarization of prompt $J/\psi$'s with 
a data that is larger than that of $\psi'$ by about a factor 100.
The prompt signal is composed of
direct $J/\psi$'s~(60\%) and $J/\psi$'s that come from decays of
the higher charmonium states $\chi_{c1}$~(15\%), $\chi_{c2}$~(15\%),  
and $\psi'$~(10\%). The contribution from the radiative decays of 
$\chi_{cJ}$'s
decreases, but does not eliminate the transverse polarization at large
transverse momentum~\cite{psipol}.
In Fig.~1(a) the theoretical prediction of the polarization parameter $\alpha$
for prompt $J/\psi$'s
is shown as a band along with the CDF data~\cite{psipol}.
While the prediction is in good agreement with the data in the moderate-$p_T$
region, it disagrees with the data in the bin at the largest $p_T$.
However, the discrepancies with the
theoretical predictions are significant only for the bin at the largest 
$p_T$, and so a definitive conclusion must await the higher statistics
measurements that will be possible in Run II of the Tevatron.
\section{Polarization of $\Upsilon(nS)$}
The CDF Collaboration also measured the polarization
of inclusive $\Upsilon(1S)$ in Run IB of the Tevatron~\cite{upspol-cdf}.
The results for the $p_T$ bins from 2 to 20 GeV and
from 8 to 20 GeV are both consistent with
no polarization. Since the cross section falls rapidly with $p_T$,
this indicates that there is little if any polarization
for $p_T$'s below about 10 GeV.
Quantitative calculations of the
polarization for inclusive $\Upsilon(nS)$ mesons
are carried out~\cite{upspol} by using ME's for direct bottomonium 
production, which have been recently determined 
from an analysis of data from Run IB at the Tevatron~\cite{B-F-L}.
There are more  feed-down channels than for prompt $J/\psi$,
but the generalization to inclusive $\Upsilon(nS)$ production is 
straightforward~\cite{upspol}.
The theoretical prediction for the polarization of 
$\Upsilon(1S)$~($\alpha=0.13\pm 0.18$ in the $p_T$ bin from 8 to 20 GeV) 
is consistent
with the recent measurement by the CDF Collaboration
in the $p_T$ bin from 8 to 20 GeV~\cite{upspol}.
It is also predicted that the transverse polarization of $\Upsilon(1S)$
should increase steadily for $p_T$ greater than about 10 GeV~(See Fig. 1(b))
and that the $\Upsilon(2S)$ and $\Upsilon(3S)$
should be even more strongly transversely polarized~\cite{upspol}.
\section{Relativistic corrections to the fragmentation process}
The CDF measurement of the polarization of prompt $J/\psi$'s
disagrees with the NRQCD prediction in the large-$p_T$ region.
Since the production rate is dominated by gluon fragmentation
in this region, it may be worthwhile to check the size of corrections that
are neglected in the available predictions for the fragmentation process.
There are many effects that could change the quantitative prediction
for $\alpha$, such as  next-to-leading order QCD corrections.
The QCD correction for the color-octet spin-triplet 
channel has been calculated~\cite{BL-nlo} but other
next-to-leading order corrections are, as yet, uncalculated.
The virtual gluon fragmentation that originates from light-quark fragmentation 
also contributes to the large transverse polarization~\cite{BL-mu}.

It is known that there are large $v^2$ corrections to various charmonium 
decay rates~\cite{BP}.~(Recently, the $v^4$ correction factors for 
$S$-wave states have been calculated~\cite{BP}.)
If there are also large $v^2$ corrections to
the gluon fragmentation into $J/\psi$, then the prediction of the
prompt $J/\psi$ may change significantly. 
The fragmentation probability is estimated by
integrating the fragmentation functions $D_n(z)$~(
$n=8$ for octet) over the longitudinal fraction $z$~\cite{v2frag}: 
\begin{eqnarray}
\int_0^1 dz \;D_8(^3S_1)(z)&\approx&(1-0.54\;v^2/0.3) 
            \int_0^1 dz \; D_8^{\rm{LO}}(z),
\end{eqnarray}
where the superscript LO denotes the limit $v\to 0$.
For charmonium, $v^2\approx 0.3$. The large negative correction~(%
$\approx 0.54$) to the color-octet spin-triplet fragmentation  
should increase the numerical value of the octet matrix element by 
about a factor $2$. The phenomenological consequences of this correction 
require further study.
\section{SUSY NRQCD}
One of the exciting new developments in quarkonium physics is its
application to low-energy supersymmetry. 
In Ref.~\cite{sb}, it was proposed that 
a light bottom squark $\tilde{b}$ with a mass similar to or less than 
the mass of the $b$ quark
may provide a solution to the puzzle that the $b$ production rate 
measured at the Tevatron is two to three times greater than the
theoretical prediction from quantum chromodynamics~\cite{bb}.
If the $\tilde{b}$ is lighter than the $b$, 
one may observe  $\tilde{b}\tilde{b}^*$
pairs in $\Upsilon$~\cite{sbsb-u} and $\chi_b$~\cite{sbsb-c} decays. 
Furthermore, forthcoming high-statistics data from the
CLEO Collaboration offer possibilities of discovery or significant 
new bounds on the existence and masses of supersymmetric particles
through the search for the monochromatic photon that is emitted from
the radiative decay of the $\Upsilon$ into $S$-wave sbottomonium~(%
a $\tilde{b}\tilde{b}^*$ bound state)~\cite{sbonium}.
\\
\\

We acknowledge enjoyable collaborations 
on the work presented here 
with Eric~Braaten~(Sec.~2-3),
\mbox{Geoffrey}~T.~Bodwin~(Sec.~4-5), and~Edmond~L.~Berger~(Sec.~5)
and we thank them for their valuable comments. 
Work in the High Energy Physics Division at Argonne National Laboratory
is supported by
the U.~S.~Department of Energy, Division of High Energy Physics, under
Contract No.~W-31-109-ENG-38.
\newpage


\end{document}